# Unveiling the Role of Friction in Coarse-Grained Clay: A Hybrid Framework Integrating Long-Range Interactions and Granular Contact Mechanics

Wang-Qi XU, Yijie WANG[1], Zhen-Yu YIN[2]

Department of Civil and Environmental Engineering, The Hong Kong Polytechnic University, Hong Kong, China

## Abstract

Given the predominant role of inter-particle physicochemical forces in governing clay behavior, researchers have increasingly utilized coarse-grained molecular dynamics (CGMD) simulations. However, inter-particle friction has been historically overlooked due to methodological limitations, and the extent to which this omission influences simulation accuracy remains an unresolved question. This study proposes a novel hybrid CGMD framework explicitly coupling long-range Buckingham potential with Hertzian granular contact mechanics. A baseline model was validated via isotropic compression, where the resulting compressibility and derived compression index ($C_c$) aligned with macroscopic geotechnical observations. Parametric analyses revealed that viscoelastic damping of particle contacts governs structural evolution. Elevated damping suppresses densification, trapping platelets in disorganized, high-void-ratio configurations. Furthermore, evaluating the interplay with thermal fluctuations underscores the necessity of precise temperature control to prevent such unphysical kinetic trapping. Finally, uniaxial compression tests demonstrate the critical importance of inter-particle friction. Explicit friction locks sliding interfaces and sustains significantly higher loads compared to frictionless models; the latter rely solely on geometric interlocking and ultimately exhibit unphysical fluid-like yielding. By bridging atomistic potentials with contact mechanics, this framework highlights the fundamental role of the inter-particle friction and offers essential guidelines for future multi-scale simulations of clay assemblies.

---

[1] Corresponding author. Email: yi-jie.wang@polyu.edu.hk

[2] Corresponding author. Email: zhenyu.yin@polyu.edu.hk

# 1. Introduction

Montmorillonite (MMT), a primary constituent of bentonite, is widely utilized in geotechnical and environmental engineering as a strategic barrier material due to its exceptionally low permeability and high swelling capacity [1, 2]. Key macroscopic mechanical properties of MMT assemblies, including stiffness, shear strength, and long-term structural stability, emerge from a complex hierarchy of multi-scale interactions. These range from molecular hydration forces and electrostatic repulsion at the nanoscale to the collective jamming and kinematic interlocking of platelets at the mesoscale. In recent years, the study of clay systems aided by advanced computer simulations has attracted growing attention. The thermodynamic behavior of MMT has been extensively investigated using molecular dynamics (MD) simulations, yielding significant insights into nanoscale swelling mechanisms [3, 4] and fundamental mechanical responses [5-7]. However, conventional all-atom MD (AAMD) is inherently constrained by its high computational cost, limiting studies to relatively small system sizes (typically below 10 nm) and exceedingly short simulation times (usually shorter than 100 ns). This orders-of-magnitude discrepancy between simulation capabilities and actual experimental timescales poses a profound challenge when attempting to directly correlate computational predictions with macroscopic geotechnical observations [8].

These constraints have motivated the development of coarse-grained MD (CGMD) approaches, which effectively simplify the system by reducing its total degrees of freedom [9]. A variety of CGMD models have successfully represented MMT as idealized ellipsoidal or hexagonal particles, replacing explicit atomic-level details with effective interactions governed by analytical potentials, such as the Gay-Berne or Morse functions [10-15]. Furthermore, systematic bottom-up CGMD methods, including Iterative Boltzmann Inversion (IBI), have been employed to bridge the scales by matching structural distribution functions derived from atomistic data [16, 17]. A primary advantage of these bottom-up strategies is that effective interaction potentials are extracted directly from atomistic trajectories, thereby bypassing the need for resource-intensive macroscopic calibration against experimental data [18, 19]. Consequently, these models remain physically grounded and offer clearer insights into the microscopic origins of macroscopic clay response. However, these traditional frameworks rely on effective potentials derived from equilibrium states, which intrinsically smooth the underlying free-energy landscape. A critical, yet often overlooked, consequence of this smoothing is the neglect of non-conservative forces, specifically inter-particle friction and localized viscous damping. It has been recently highlighted that the absence of friction in CGMD leads to unphysical behavior and an inability to capture realistic jamming transitions [20]. Despite these warnings, it remains unclear to what extent these ignored parameters govern the resulting clay fabric and mechanical response, as no framework currently exists to systematically investigate their influence within an MD environment.

In the discrete element method (DEM), frictional contact dissipation is standard for capturing the macroscopic strength and consolidation of granular assemblies [21, 22]. Although recent advancements have extended DEM to clay modeling by incorporating van der Waals (vdW) interactions and electrostatic double-layer (EDL) repulsion [23, 24], these frameworks typically rely on macroscopic calibration, which is a process that necessitates numerous empirical assumptions and can introduce significant subjectivity. Consequently, the fundamental question of how inter-particle friction governs clay behavior remains unresolved, as no existing tool can simultaneously capture both contact dissipation and atomistically-derived physics.

To bridge this gap, this study proposes a pioneering hybrid CGMD framework that seamlessly integrates granular contact mechanics within a bottom-up molecular environment, thereby reducing the reliance on empirical assumptions while maintaining physical fidelity across scales. By integrating an explicit Hertzian formulation with long-range Buckingham potentials, and calibrating the system against the foundational model of Zhang et al. [13], this approach robustly accounts for inter-particle friction and damping, which are often neglected in conventional MD potentials. Initially, a baseline model is defined and validated through incremental isotropic compression. The resulting compressibility confirms that the framework accurately captures the characteristic mechanical response of clay assemblies, establishing a physically sound foundation for subsequent investigations. A comprehensive parametric analysis is then performed to quantify how varying normal and tangential damping governs compression kinetics and fabric evolution. Furthermore, the interplay between mechanical damping and thermal fluctuations is examined to demonstrate the critical role of temperature control in preventing unphysical kinetic trapping. Finally, uniaxial compression tests are utilized to prove the indispensability of explicit inter-particle friction in driving shear strength. Ultimately, this hybrid framework provides a vital link between micro-scale contact physics and the macroscopic mechanical reality of clay assemblies.

# 2. Methodology

## 2.1 Interactions between Clay Particles

The interaction between clay particles is governed by a complex interplay of forces that vary significantly with the separation distance, as illustrated in Figure 1. Conventional theoretical calculations usually focus on vdW and EDL effects, as described by the Derjaguin–Landau–Verway–Overbeek (DLVO) theory. However, at separations below a critical threshold, the discrete nature of water molecules and exchangeable cations gives rise to powerful hydration forces [25]. These short-range forces often dominate the mechanical response, providing hydration-mediated resistance. Crucially, inter-particle friction can be activated within these hydration layers prior to true steric contact; thus, significant shear resistance is observed even when the mineral surfaces are not in direct atomic contact [26].

To capture this multi-regime physics, a dual-component approach is adopted in the modelling strategy. For the long-range regime, a Buckingham potential is employed to represent the combined effect of vdW and EDL interactions:

$$E_{Buckingham} = Ae^{-r/\rho} - \frac{C}{r^6} \tag{1}$$

where $A$ and $C$ are coefficients for the repulsive and attractive energy, $\rho$ is an ionic-pair dependent length parameter and $r$ is the separation distance between two particles. For the short-range regime, a Hertzian force field with a viscoelastic damping style is employed to represent the hydration-dominated state. The normal component of the contact force is given by

$$\mathbf{F}_n = k_n R_{eff}^{1/2} \delta_{ij}^{3/2} \mathbf{n} - \eta_n \mathbf{v}_n \tag{2}$$

where $k_n$ is spring stiffness, $R_{eff}$ is the effective radius, $\delta_{ij}$ is the particle overlap (representing the compression of the hydration shell), $\eta_n$ is the normal damping coefficient, and $\mathbf{v}_n$ is the normal component of relative velocity. The tangential force is presented as

$$\mathbf{F}_t = -\min\left(\mu\|\mathbf{F}_n\|, \left\|x_{\gamma,t}\eta_n\mathbf{v}_t\right\|\right)\mathbf{t} \tag{3}$$

where $\mu$ is the friction coefficient, $x_{\gamma,t}$ is the scaling coefficient, and $\mathbf{v}_t$ is the tangential component of relative velocity. By allowing a controlled degree of particle overlap ($\delta_{ij}$), the framework mimics the compressibility of hydration layers while explicitly incorporating the tangential friction and viscous damping that govern energy dissipation during particle sliding and jamming.

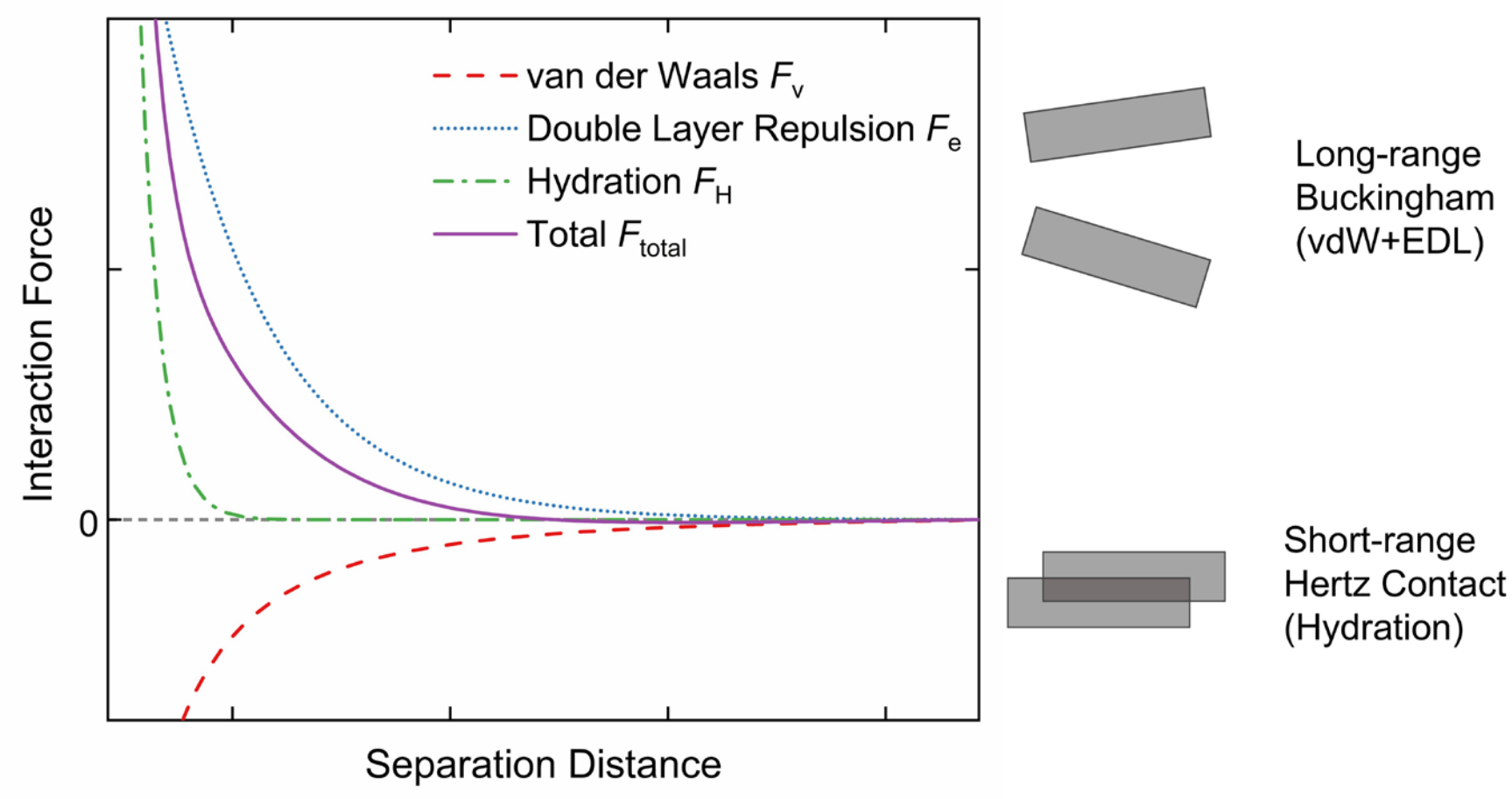

Figure 1 Schematic representation of the inter-particle force-distance relationship and the dual-component modeling strategy. Grey rectangles denote clay platelets under saturated conditions.

## 2.2 Platelet Geometry and Energy Fitting

The CG-MMT model is constructed as a hexagonal platelet with a diameter of 120 Å, consistent with the mesoscale geometry proposed by Zhang et al. [13]. As shown in Figure 2, a single platelet is composed of 271 overlapping spheres, each with a diameter of 8.5 Å. To ensure structural integrity and prevent unphysical inter-platelet penetration, the spheres are arranged to overlap, with the center-of-mass (COM) distance between adjacent spheres fixed at 5.5 Å. The constituent spheres are categorized into inner (red) and outer (blue) groups based on their spatial positions. While they share identical diameters and masses, they are assigned distinct interaction parameters to accurately capture edge effects. The mass of each sphere is calibrated to reflect an 80% hydrated state. Within the simulation framework, a hybrid atom style (sphere and molecular) is employed, enabling each platelet to be treated as a single rigid molecule during the computation.

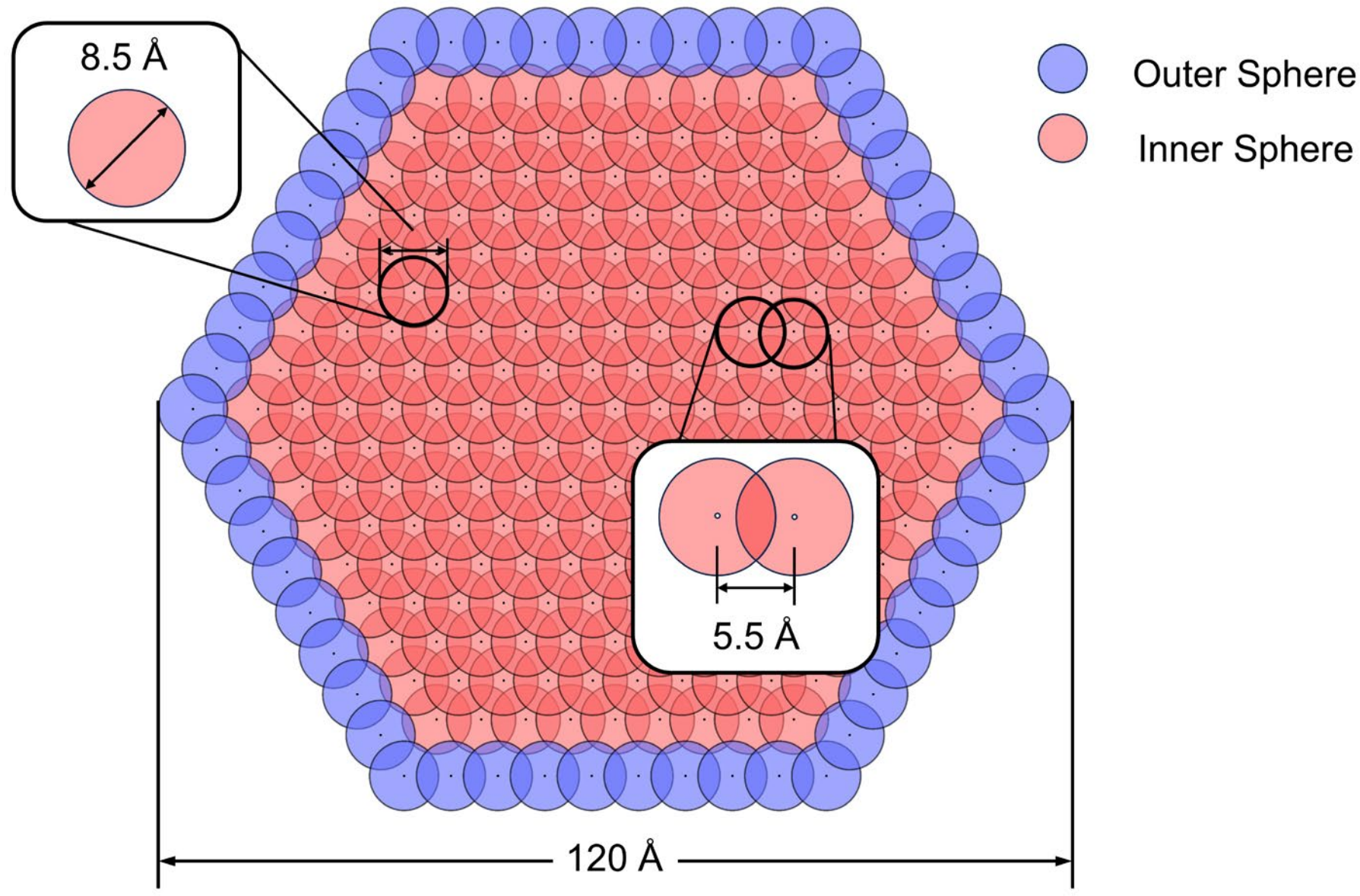

Figure 2 Schematic of the CG-MMT platelet, modeled as two types of equal-sized, overlapping spheres.

The total potential energy between an interacting pair is defined as the sum of the Buckingham term and the integrated Hertz energy as

$$E_{total} = E_{Buckingham} + \int_0^{\delta} F_{Hertz}(\delta)d\delta \quad (4).$$

Within this formulation, the Hertzian term dominates at short ranges to enforce hard-sphere-like contact and prevent unphysical particle overlaps. This combined model is subsequently used to fit the reference energy curves proposed by Zhang et al. [13] Given the high dimensionality of the parameter space, the calibration is conducted through a systematic, stepwise approach. Initially, face-to-face and edge-to-edge interactions between two platelets are fitted independently to determine the interaction parameters for the inner-inner and outer-outer sphere configurations, respectively (Figure 3(a)(b)). Subsequently, the cross-interaction parameters for the inner-outer spheres are tuned using face-to-edge interactions, as shown in Figure 3(c).

During this calibration phase, the potential energy surfaces are obtained from a series of steady-state configurations. Consequently, the fitting focuses exclusively on the conservative force components. This allows for the determination of the Buckingham potential parameters and the normal spring stiffness ($k_n$) of the Hertzian model, while the dissipative damping coefficients are reserved for independent investigation, detailed in the following sections. The resulting optimized parameters are summarized in Table 1, and the derived force-distance curves are used to validate the mechanical consistency of the fitted potentials against the reference data (Figure 4). Note that the $C$ parameter in the Buckingham potential is set to zero due to the absence of significant attractive interactions. Although some attraction exists in edge-to-edge cases, its magnitude is comparatively small.

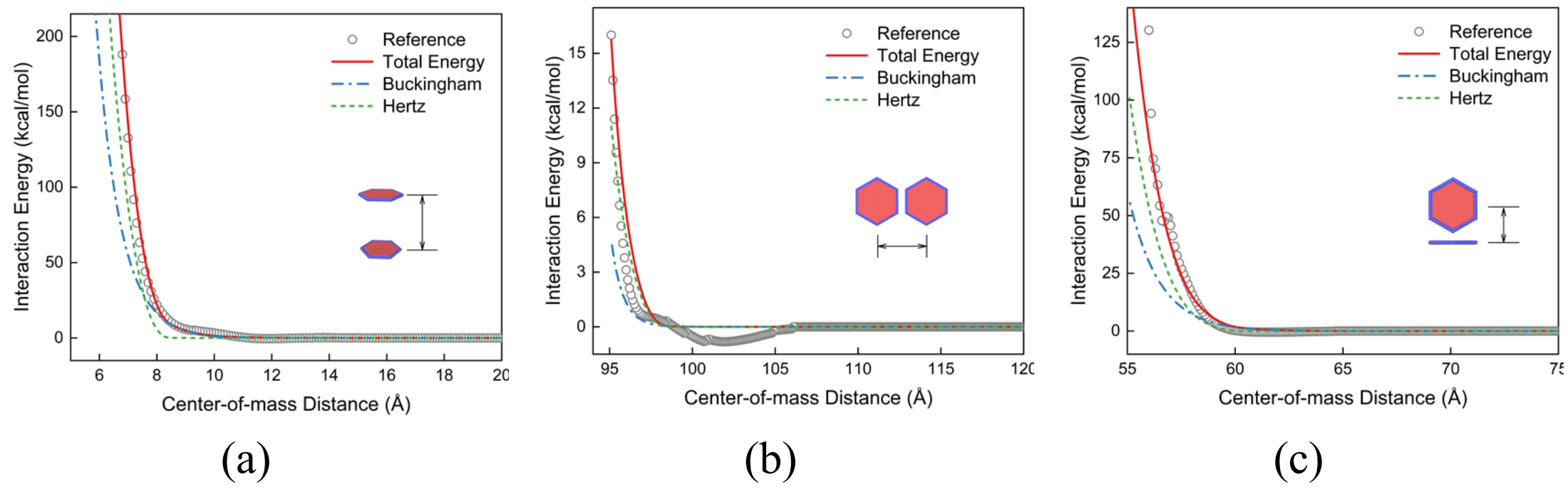

Figure 3. Calibration of the hybrid potential (Buckingham + Hertz) through optimized fitting to MD reference energy profiles for (a) face-to-face, (b) edge-to-edge and (c) face-to-edge interactions.

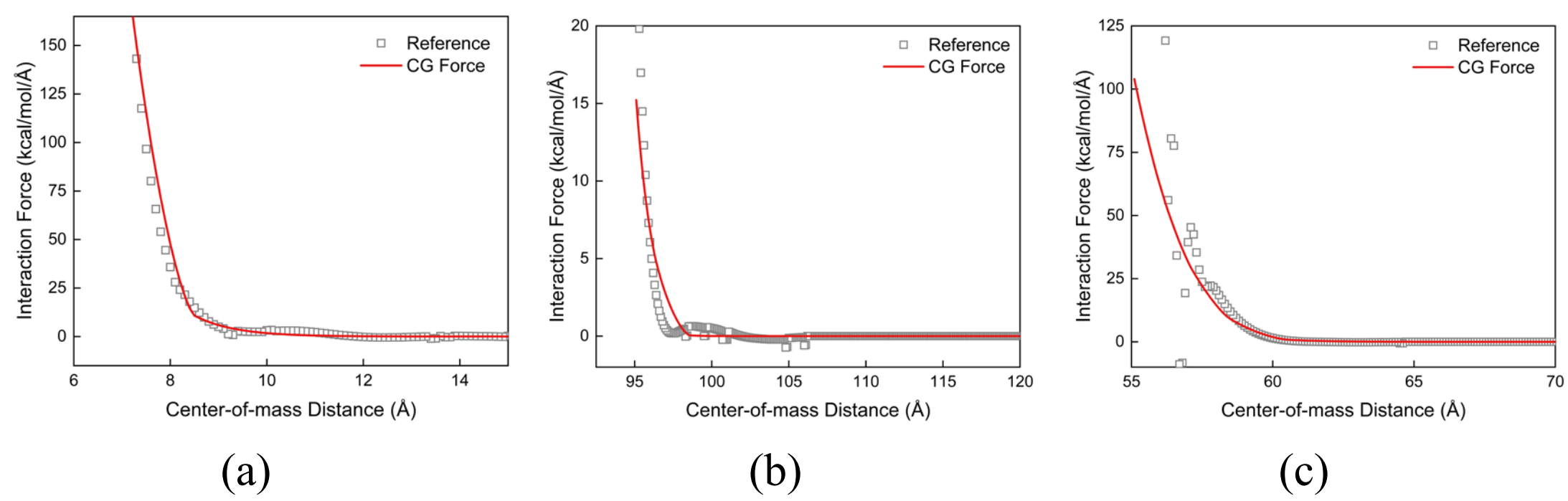

Figure 4 Force-distance profiles for the CG-MMT (a) face-to-face, (b) edge-to-edge and (c) face-to-edge interactions against benchmark MD results.

## 2.3 Isotropic and Uniaxial Compression

All MD simulations in this study were performed using the Large-scale Atomic/Molecular Massively Parallel Simulator (LAMMPS) code [27], with inter-particle contact mechanics implemented via the LAMMPS granular package [28]. The initial system was prepared by placing 1000 MMT platelets in a sparse, uniform configuration with random orientations inside a cubic box measuring 1400 Å on each side, as illustrated in Figure 5. The use of 1000 platelets represents a reasonable compromise between computational feasibility and the statistical representation of microstructural evolution [10, 29]. This system size is consistent with recent CG simulations of clay systems, which have employed assemblies ranging from several tens to several thousand clay platelets [17, 30]. To optimize computational efficiency, each clay platelet was treated as a rigid body, thereby neglecting intra-platelet pairwise interactions. An integration timestep of 5 fs was utilized for all simulations. To model the assembly of clay minerals at room temperature, the system was first equilibrated in the canonical (NVT) ensemble at 300 K for 50 ns. Initial velocities were randomly assigned from a Gaussian distribution, and a Nosé-Hoover thermostat was applied to regulate both the translational and rotational degrees of freedom for the rigid bodies. Following this thermal equilibration, the CG-MMT system was isotropically compressed at an initial pressure of 1 atm for 10 ns, followed by a subsequent compression stage at 10 atm. The system pressure was maintained using the *deform/pressure* command, which dynamically adjusted the simulation box dimensions to achieve the target stress state.

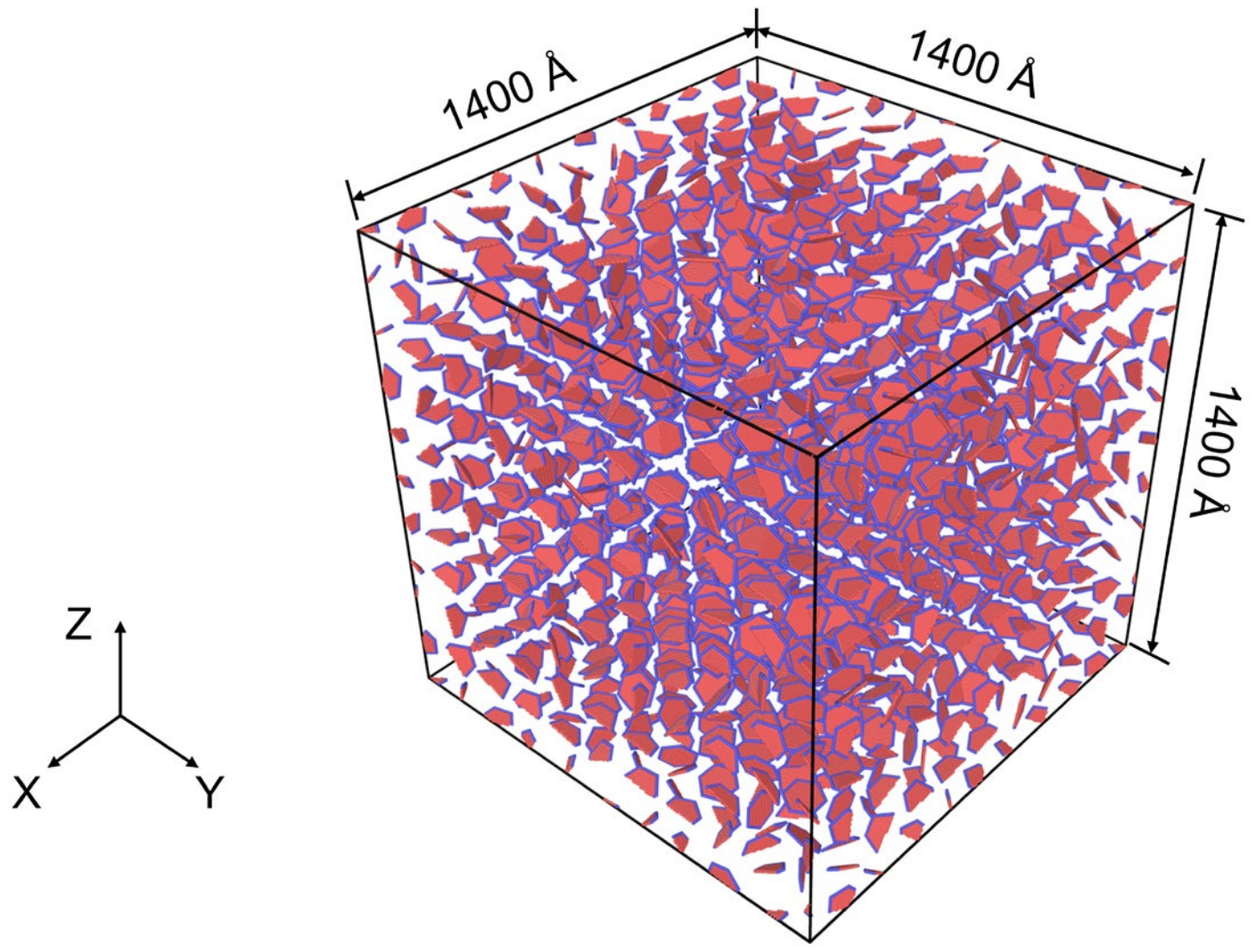

Figure 5 Snapshot showing the initial configuration of the CG-MMT model containing 1000 platelets.

To isolate the role of mechanical energy dissipation, an alternative isotropic compression protocol was conducted for purely friction-driven assemblies. By adopting the microcanonical (NVE) ensemble without temperature control for the same initial configuration, all kinetic energy was dissipated exclusively through the prescribed normal and tangential damping forces, rather than a thermostat.

For both protocols, the system was considered fully isotropic compressed under 10 atm once the thermodynamic parameters reached a steady state. This was defined as the point where the absolute slope of the evolution curve for these quantities fell below a threshold, which was set as 0.01% per nanosecond in this study. The equilibrated structures served as the initial configurations for subsequent mechanical testing. Uniaxial compression tests along the $z$-axis under laterally confined conditions at 300 K were performed. A constant axial strain rate of $10^7$/s was applied. While high by macroscopic experimental standards, this represents a relatively slow rate for MD time scales, ensuring a balance between computational feasibility and physical accuracy [31, 32]. No strain is allowed in the $x$- or $y$- directions. Stress components and system configurations were sampled continuously throughout the deformation process to characterize the mechanical response.

# 3. Results and Discussion

## 3.1 Model Validation

To investigate the influence of friction parameters, a baseline model was first established. The parameters were selected based on physical and numerical considerations. The friction coefficient ($\mu$) was set to 0.1, which falls within the reported range for saturated clay minerals, consistent with both simulations and experiments [33-35]. The normal damping coefficient ($\eta_n$) of 0.2 was chosen to provide sub-critical energy dissipation, allowing for realistic structural rearrangements. The selection of the value was informed by the theoretical critical damping coefficient $\eta_{crit}$, defined for a linear spring-dashpot system as [36]

$$\eta_{crit} = 2\sqrt{m_{eff}k_{eff}} \quad (5).$$

Based on the particle mass and contact stiffness of the CG-MMT model, the estimated critical thresholds for the various interaction types in this study range from approximately 0.18 to 0.73. Consequently, $\eta_n$ was selected as 0.2 to provide sufficient energy dissipation for a stable quasi-static assembly. The tangential scaling factor ($x_{\gamma,t,}$) was set to 5.0.

To validate the macroscopic mechanical consistency of this baseline, an incremental isotropic compression was performed at pressures of 1, 3, 10 and 100 atm (corresponding to approximately 100, 300, 1000, and 10000 kPa). The resulting void ratio-pressure ($e$-log $p$) relationship is compared against experimental benchmarks for smectite, bentonite and clayey soil in Figure 6. The simulation captures a distinct bimodal compression response. In the low-to-mid pressure range ($\leqslant$ 1000 kPa), the compression index ($C_c$) calculated is 2.17, aligning well with the typical range of 1.0 to 2.6 reported for bentonites [37]. When the pressure is high (> 1000 kPa), the assembly exhibits a significantly stiffer response with a $C_c$ of 0.39, accurately reflecting the reduced compressibility as platelets reach maximum compaction. The strong agreement between the simulated compression path and the experimental data [37-40] confirms that the hybrid CG-MMT model captures the fundamental macroscopic behavior of clay. This establishes a physically sound foundation for the subsequent investigations into the hidden influence of dissipative parameters.

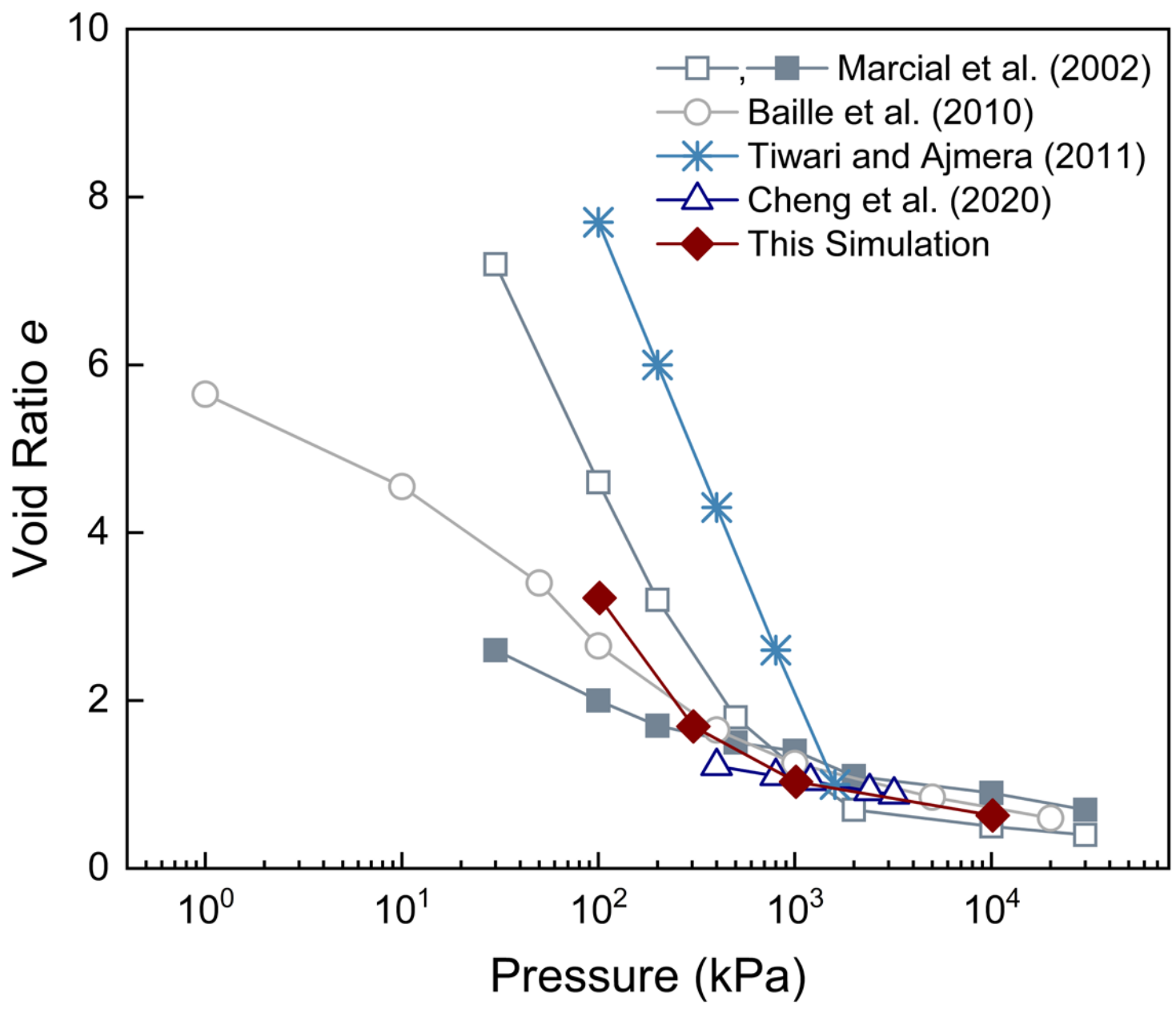

Figure 6 Compressibility of the hybrid CG-MMT model compared against experimental results on smectite from Marcial et al. (2002), bentonite from Baille et al. (2010), MMT from Tiwari and Ajmera (2011) and clayey soil from Cheng et al. (2020).

## 3.2 Influence of Viscoelastic Damping Coefficients

Following the successful validation of the baseline model, this section investigates the governing role of the viscoelastic damping parameters, $\eta_n$ and $x_{\gamma,t}$, which are critical for representing the inelastic energy dissipation occurring during particle collisions. While these parameters are often neglected in traditional conservative CGMD frameworks, they represent a fundamental component of the mesoscale physics of clay. In the absence of direct experimental data for these specific CG interactions, a systematic parametric sensitivity analysis was conducted to evaluate how energy dissipation rates influence the resulting clay assemblies. To ensure a rigorous comparison, all simulations were initialized from an identical spatial configuration, and the specific damping coefficients were applied consistently throughout the entire simulation sequence, including the initial NVT equilibration and the subsequent isotropic compression. This approach allows for a comprehensive assessment of how energy dissipation dictates the system's structural trajectory, moving from an initial random state to its final thermodynamic and mechanical equilibrium.

### 3.2.1 Sensitivity to Normal Damping

To evaluate the influence of energy dissipation on the evolution of the clay fabric, a parametric study was conducted using three normal damping coefficients: $\eta_n = 0.2$, 1.0, and 2.0. In contrast to the baseline model, $\eta_n = 1.0$ and 2.0 represent overdamped regimes, which were tested to observe the effects of rapid numerical quenching on the final equilibrium configuration.  For these sensitivity tests, the tangential damping scaling parameter, $x_{\gamma,t}$, was held constant at 5.0 and friction coefficient ($\mu$) was set as 0.1.

The systems were deemed equilibrated under 10 atm pressure once the evolution of system volume met the stability criteria defined in Section 2.2 (i.e., a slope threshold below 0.01% per nanosecond). As shown in Figure 7, the time required to reach equilibrium varied significantly across the three regimes. Cases with lower damping coefficients ($\eta_n = 0.2$ and 1.0) converged to the target pressure within 25 ns. In contrast, the heavily overdamped case ($\eta_n = 2.0$) exhibited a more sluggish response, requiring approximately 50 ns to satisfy the equilibration criteria.

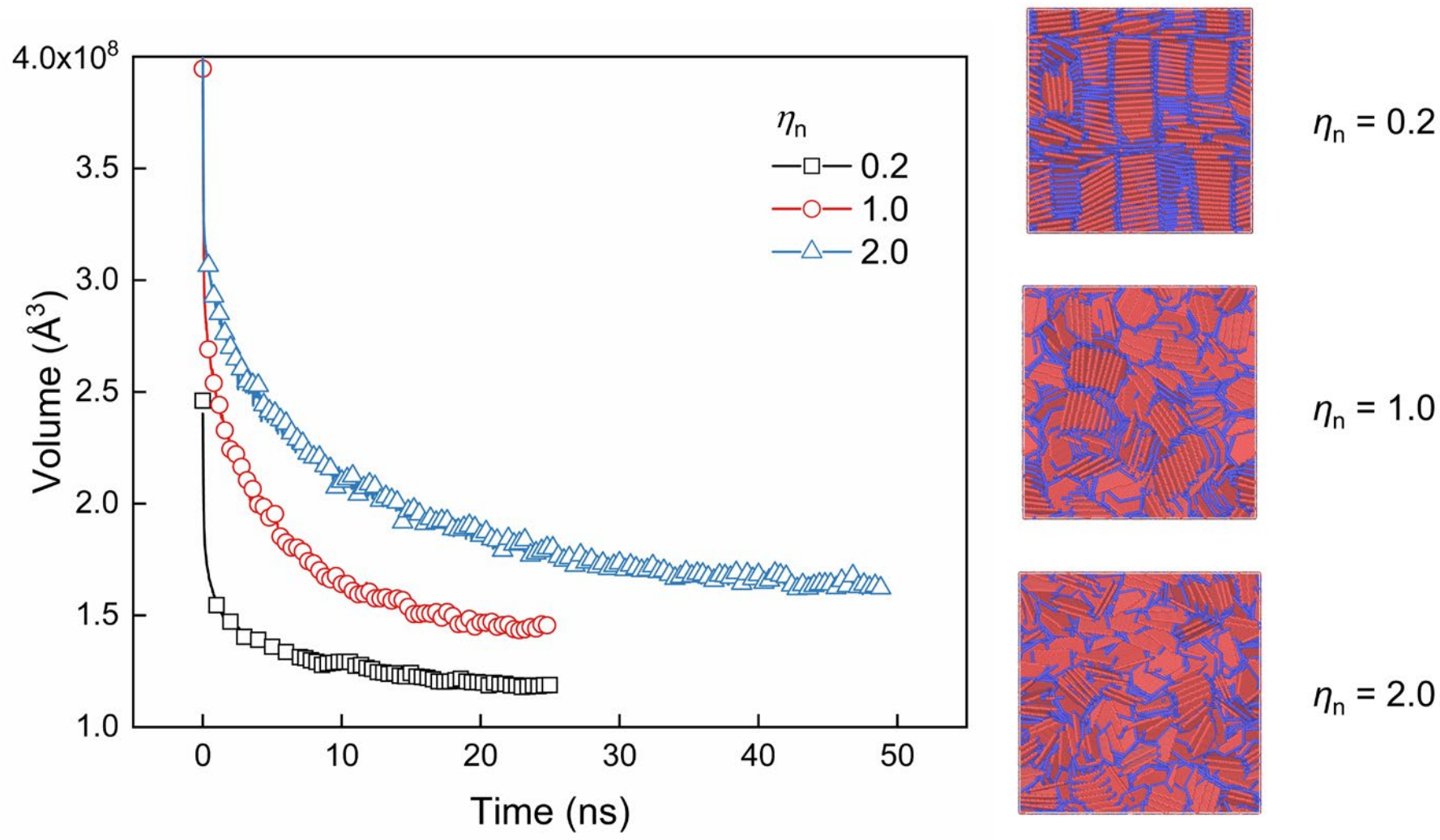

Figure 7 Effect of normal damping coefficients ($\eta_n$) on system volumetric evolution during isotropic compression and post-compression equilibrium configurations (plan view perpendicular to the *y*-axis)

Following isotropic compression, the resulting microstructures were analyzed to quantify the influence of energy dissipation on the clay fabric. Specifically, the void ratio ($e$) and the orientational order parameter ($S$) were calculated to characterize the packing density and the degree of platelet alignment, respectively. The order parameter S provides a scalar measure of how closely the platelets align with a common axis, known as the system director (**n**). $S$ is defined as [41, 42]

$$S = \langle \frac{3cos^2\theta - 1}{2} \rangle \tag{6}$$

where $\theta$ is the angle between the normal vector of an individual platelet (**u**) and the system director **(n)** [43]. The director **n** represents the preferred orientation axis of the assembly and is identified as the eigenvector corresponding to the largest absolute eigenvalue of the second-order orientation tensor, $q_{ij}$

$$q_{ij} = \frac{1}{N} \sum_{m=1}^{N} \left( u_i u_j - \frac{1}{3} \delta_{ij} \right) \tag{7}$$

where $N$ denotes the total number of platelets, and $\delta_{ij}$ is the Kronecker delta. The resulting value of $S$ ranges from 0, indicating a perfectly isotropic and random distribution, to 1, representing a state of perfect parallel alignment.

Table 2 summarizes the structural properties, while Figure 7 provides snapshots of the configurations after isotropic compression for visual inspection. A clear trend emerges: an increasing normal damping coefficient ($\eta_n$) significantly hinders structural densification. This results in a substantially higher void ratio, with $e$ nearly doubling from 1.03 to 1.97 as $\eta_n$ increases from 0.2 to 2.0. Additionally, high normal damping suppresses system alignment, evidenced by the sharp decline in the order parameter from 0.86 to 0.04. These results suggest that elevated damping freezes the initial random configuration, effectively preventing the transition into ordered domains.

### 3.2.2 Sensitivity to Tangential Damping

To isolate the specific role of tangential resistance in governing shear stability and energy dissipation, a decoupled parametric study was employed. In this phase of the investigation, the normal damping coefficient ($\eta_n$) was fixed at the validated baseline value of 0.2 to ensure the system achieved a stable, converged equilibrium state, while the tangential damping was varied to characterize its influence on the sliding and rotational mobility of the platelets. Specifically, the tangential scaling factor ($x_{\gamma,t}$) was set to 1, 5 (baseline), and 10, with the friction coefficient ($\mu$) held constant at 0.1. All other simulation parameters remained consistent across cases to ensure that any observed structural variations were purely a function of tangential resistance. This systematic isolation allows for a precise assessment of how tangential dissipation governs the internal rearrangement and eventual alignment of the clay fabric.

The resulting configurations exhibit a high sensitivity to tangential damping, as quantitatively summarized in Table 2 and visualized in Figure 8. Lower values of $x_{\gamma,t}$ facilitate significant structural rearrangement and strong local alignment. Specifically, models with $x_{\gamma,t} = 1$ and 5 both achieve high order parameters ($S > 0.8$) and low void ratio (1.03 and 1.06), suggesting that reduced tangential resistance allows platelets to slide and rotate into a dense, ordered packing. However, despite their similar degrees of order, these two cases converge to markedly different system directors (**n**). At $x_{\gamma,t} =$ 1, the preferred orientation is aligned almost entirely with the $x$-axis ($\mathbf{n} \approx [-1.0, 0, 0]$), whereas at $x_{\gamma,t} = 5$, the alignment shifts to the $z$-axis ($\mathbf{n} \approx [0, 0, 1]$).

This variation indicates that while low tangential damping facilitates structural ordering,

the ultimate orientation of the resulting domains remains inherently stochastic and highly sensitive to subtle fluctuations in energy dissipation during isotropic compression. This behavior is fundamentally tied to the high aspect ratio of the CG-MMT platelets; unlike idealized spherical assemblies, the extreme anisotropy of the platelets means that kinetic agitation readily triggers spatial inhomogeneities and localized alignment [44]. Notably, in a significantly larger system, these localized domains would likely orient in diverse directions and average out, potentially leading to a more globally isotropic fabric under isotropic compression. Nevertheless, the primary objective here is to reveal the mechanistic trends introduced by inter-particle friction. While computational resource constraints currently limit the exploration of larger domains, the observed transition from ordered to disordered states remains qualitatively robust, as the role of tangential resistance in restricting particle mobility is a physical mechanism independent of the total particle count. In contrast, increasing the tangential damping to $x_{\gamma,t}$ = 10 significantly inhibits particle rotation and relative sliding. This is evidenced by a substantial drop in the order parameter to $S$ = 0.37 and an increase in the void ratio to 1.23. Under these conditions, the elevated tangential resistance rapidly dissipates the kinetic agitation required for local rearrangement, effectively locking the particles in a disorganized, isotropic-like state and preventing the system from reaching the more compact, ordered equilibrium observed at lower damping levels.

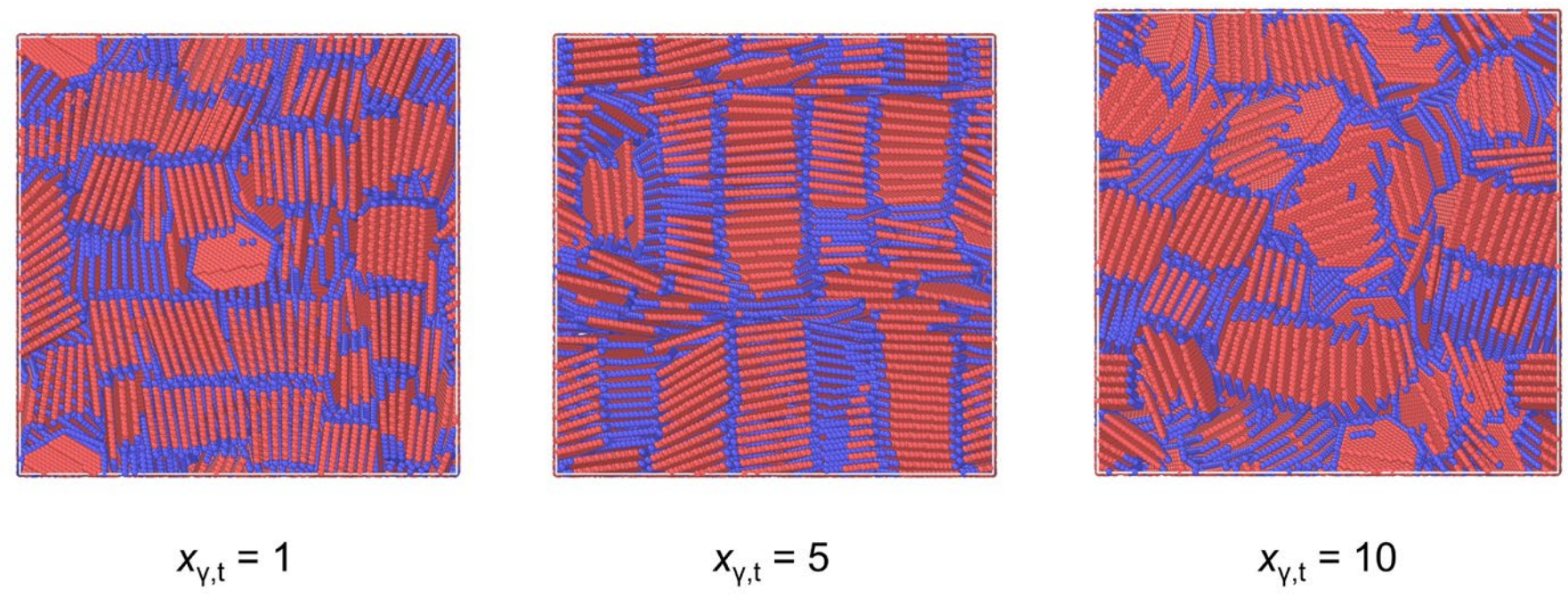

Figure 8 Plan view (perpendicular to the $y$-axis) of the post-compression configurations of models with various tangential damping scaling factors ($x_{\gamma,t}$).

## 3.3 Effect of Temperature Control

To explore the effects of different energy damping mechanisms, an identical initial configuration of 1000 randomly positioned platelets was isotropically compressed without a thermostat. The damping parameters for the contact model were set to $\eta_n$ = 0.2, $x_{\gamma,t}$ = 5, $\mu$ = 1.0 in this trial. The absence of a thermostat during compression means that kinetic energy dissipated solely through the damping force. Isotropic compression for this case completed in approximately 25 ns, comparable to the system utilizing the Nosé-Hoover thermostat. However, the resulting microstructure differed drastically. The unthermostatted configuration was significantly more isotropic and less dense, demonstrated by a sharp drop in the order parameter ($S$ = 0.09, down from 0.86) and an increased void ratio ($e$ = 1.35, up from 1.03).

The absence of a thermostat, combined with a strict reliance on normal and tangential damping, fundamentally alters the resting state of MMT platelets by shifting the system from thermodynamic equilibrium to a purely mechanical, athermal state. In a standard thermostatted environment, thermal fluctuations provide the platelets with continuous kinetic energy (maintained around 1800 kcal/mol as shown in Figure 9), allowing them to overcome small potential barriers and explore a dynamic, free-energy-minimized states, such as forming reversible face-to-face tactoids or edge-to-face networks. However, when the thermostat is removed, there is no mechanism to re-inject kinetic energy into the system. Normal and tangential damping forces continuously drain energy during platelet collisions and sliding interactions, effectively acting as an aggressive quenching process that drops the kinetic energy to near-zero. Consequently, the platelets rapidly lose their mobility and arrest into a static, frozen configuration. This forced zero-Kelvin state traps the MMT platelets in local, metastable energy minima based entirely on their collision history and contact mechanics, preventing them from reorganizing into their true, thermally stable resting state.

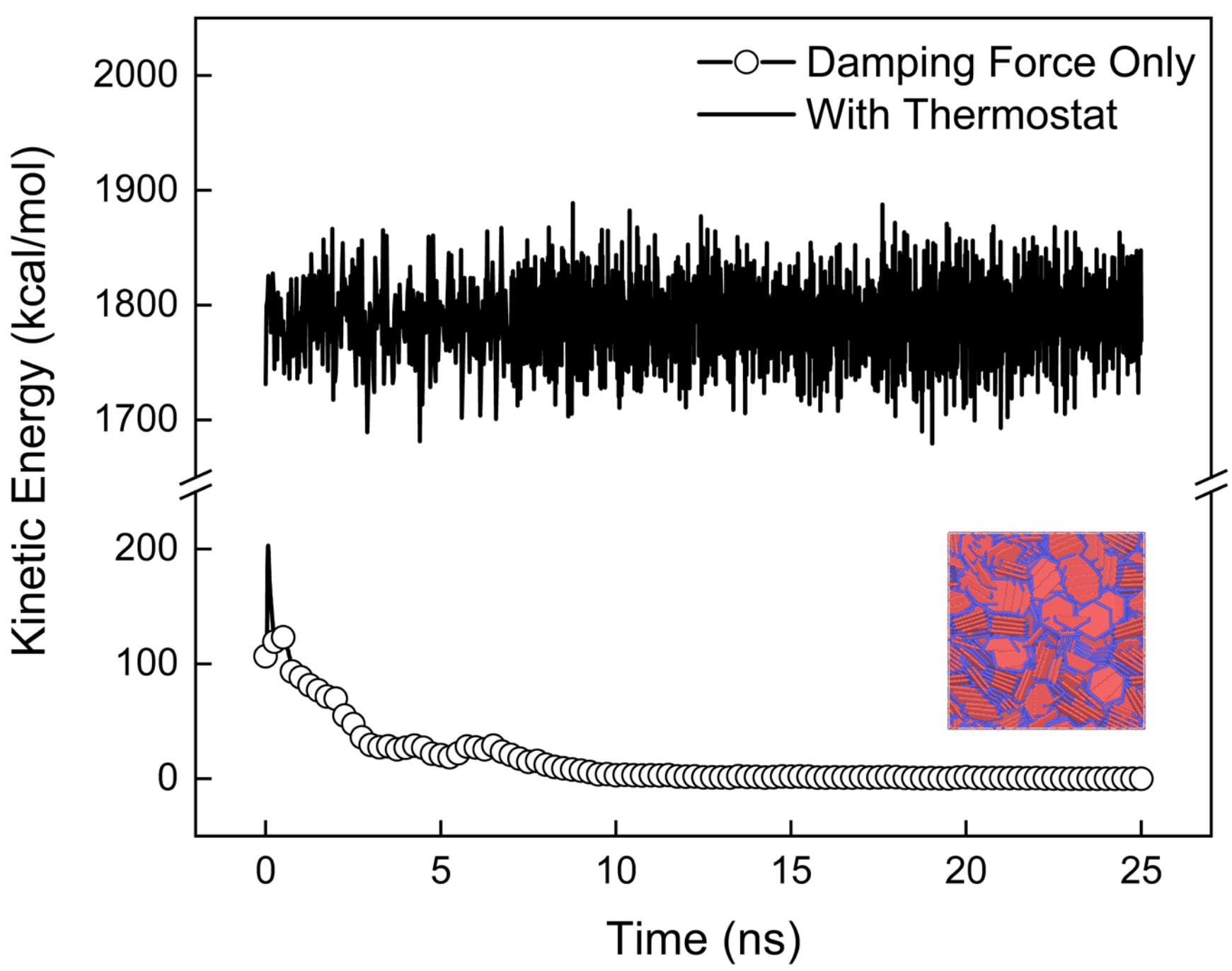

Figure 9 Influence of temperature control on kinetic energy dissipation during isotropic compression of CG-MMT. The inset displays the final configuration for the damping-only model.

In summary, while a thermostat is essential for simulating natural self-assembly and preventing artificial kinetic trapping, a purely damping-driven approach is optimal for rapidly quenching systems to local mechanical minima or isolating athermal contact mechanics. Previous research highlights that while thermostats are necessary for investigating temperature-dependent phenomena, they can also artificially distort structural dynamics and relaxation [45, 46]. Ultimately, the chosen energy dissipation mechanism should be precisely tailored to the specific physical behaviors under investigation.

## 3.4 Role of Inter-Particle Friction Coefficient

To characterize the mechanical response of the validated baseline CG-MMT assembly, uniaxial compression tests were performed under laterally confined conditions on the baseline model ($\eta_n = 0.2$, $x_{\gamma,t} = 5$, $\mu = 0.1$). The resulting axial stress-strain behavior is depicted in Figure 10. The initial response, up to approximately 10% axial strain, exhibits a relatively linear increase in stress. A linear fit of this region yields an estimated compressive modulus ($E_s$) of 8.08 MPa ($R^2 = 0.90$), representing the elastic resistance of the dense, ordered clay fabric. While this modulus is significantly lower than the GPa-scale values typically derived from AAMD simulations, [47-50], it aligns closely with macroscopic experimental measurements for clay, which generally fall in the range of several to several tens of MPa [51-53]. This scale-dependent contrast is fundamentally expected. AAMD primarily captures the stiffness of the rigid crystalline lattice of individual clay mineral, whereas the proposed CG-MMT model captures the macroscopic compliance of the bulk assembly, which is governed by inter-platelet sliding, void dynamics, and frictional contacts.

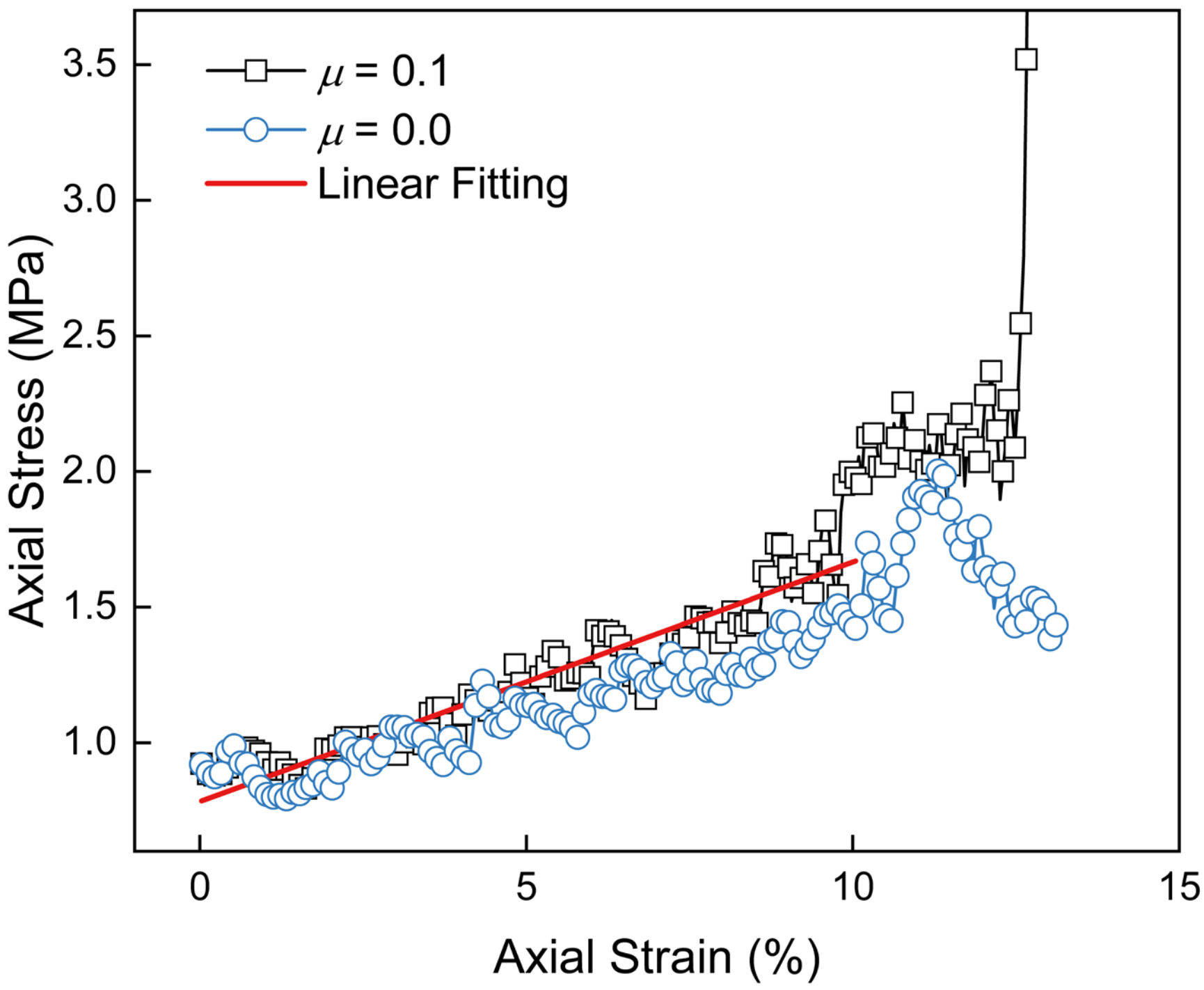

Figure 10 Stress-strain curve of the uniaxial compression for frictional and frictionless models.

Beyond 10% strain, the baseline system undergoes a dramatic transition characterized by a sharp, non-linear increase in axial stress. This strain-hardening effect likely stems from the significant reduction in void ratio, where the particles reach a state of maximum compaction and the repulsive components of the potential dominate the response. The underlying structural mechanisms are tracked in Figure 11. The void ratio ($e$) decreases linearly from an initial 1.03 to 0.74 at 15% strain, confirming continuous densification throughout the loading process. Interestingly, the order parameter ($S$) remains remarkably stable during the early loading phase, even increasing slightly to S

≈ 0.88 at 5% strain. This indicates that the initial uniaxial compressive load tightens the parallel alignment established during isotropic compression. However, as the system approaches the high-stress regime (> 10% strain), $S$ begins to decline, dropping to 0.75 by the end of the simulation. This suggests that while the system is becoming denser, the extreme axial pressures eventually force local rotations or buckling of the platelets, slightly disrupting the highly ordered fabric.

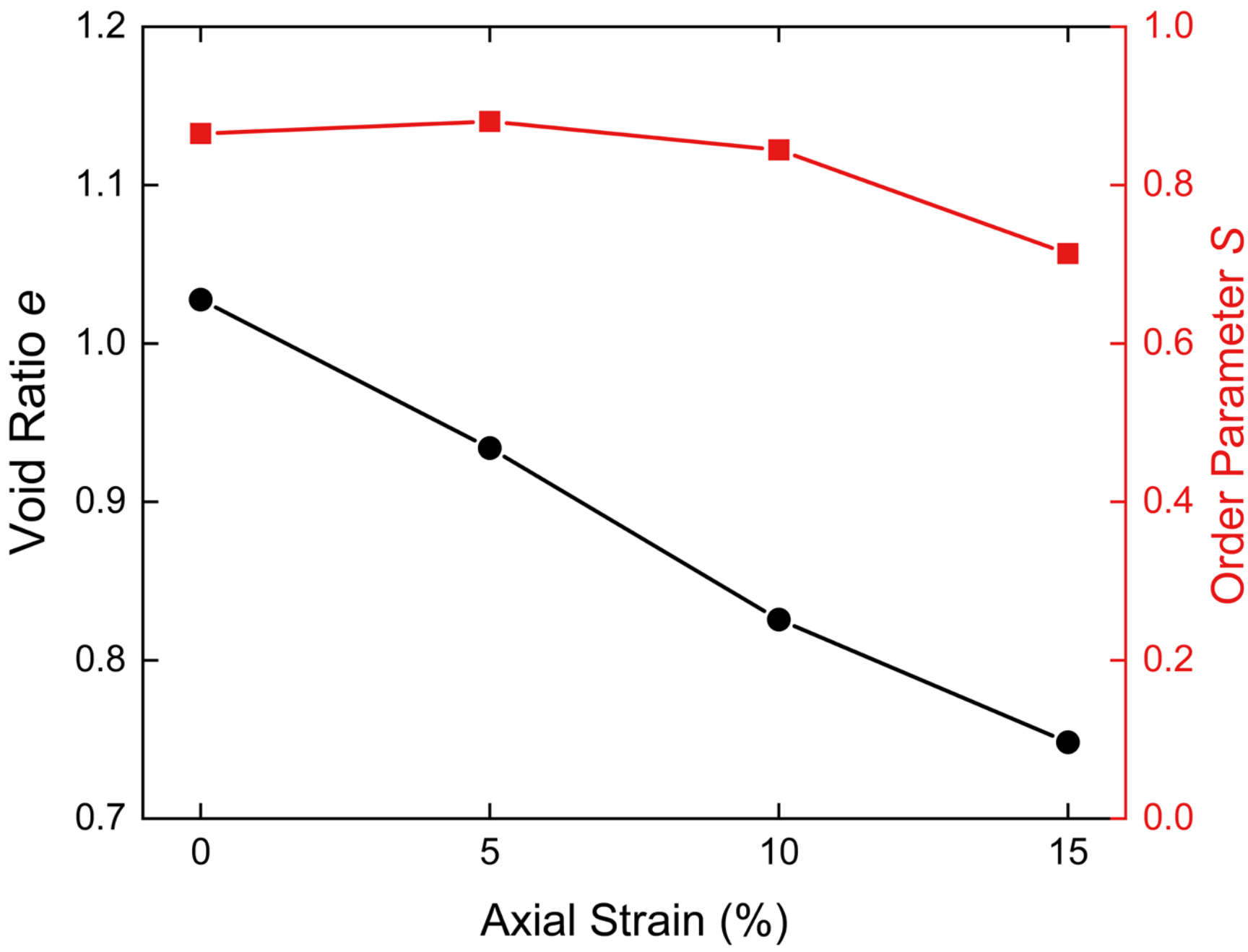

Figure 11 Evolution of structural properties of the baseline model during uniaxial compression.

To isolate the specific contribution of friction, the response of the baseline was compared against a frictionless control mode ($\mu$ = 0), as shown in Figure 10. Notably, for the frictionless case, $\mu$ was set to 0 only at the onset of the compression test. Maintaining $\mu$ = 0.1 during the prior isotropic compression ensured that both the frictional and frictionless simulations commenced from a perfectly consistent initial configuration. While both models exhibit comparable initial moduli, a significant divergence occurs beyond 10% strain. In the baseline model ($\mu$ = 0.1), the tangential resistance at contact points effectively locks the sliding interfaces, allowing the assembly to sustain much higher axial loads. Conversely, in the frictionless case, the absence of shear resistance allows platelets to slide past one another with minimal energy cost. This leads to a much lower peak stress (~ 2 MPa) followed by a noticeable strain-softening response, as the resistance in the frictionless system is derived solely from geometric interlocking. This divergence provides strong evidence of the necessity for explicitly incorporating friction in CG models to prevent the unphysical fluid-like behavior observed in traditional conservative frameworks and to accurately capture the structural integrity of clay assemblies.

## 4. Conclusions

This study developed a hybrid CGMD framework to investigate the often-overlooked influence of dissipative parameters on montmorillonite assemblies. By integrating granular contact mechanics with molecular potentials and validating the baseline through isotropic compression, the following conclusions are drawn:

1. Friction parameters control the physical realism of clay fabric: The viscoelastic damping parameters govern the structural evolution of the clay assembly during isotropic compression. High normal damping ($\eta_n$) severely hinders densification and suppresses alignment, nearly doubling the void ratio ($e$) and causing a sharp decline in the order parameter ($S$). Similarly, elevated tangential damping ($x_{\gamma,t}$) locks platelets into disorganized, isotropic-like states by inhibiting particle rotation and sliding. Lower tangential resistance is required to facilitate the rearrangement into dense, highly ordered packings. For future clay CGMD modeling, it is recommended to treat the damping parameters carefully to prevent the formation of artificially locked clay fabrics.
2. Thermal fluctuations are required for structural equilibrium: Compressing the system isotropically without a thermostat demonstrates that purely damping-driven dissipation rapidly quenches the assembly into a static, frozen configuration locked in local, metastable energy minima. A thermostatted environment is essential for providing the continuous kinetic energy necessary for platelets to overcome small potential barriers and self-assemble into true, free-energy-minimized resting states.
3. Friction drives shear strength and stability: Uniaxial compression tests reveal a critical mechanical transition at approximately 10% axial strain. While both frictional and frictionless models show similar initial elastic responses, the presence of inter-particle friction ($\mu = 0.1$) allows the assembly to sustain significantly higher axial loads by locking sliding interfaces. Conversely, frictionless assemblies suffer from structural yielding, lower peak stress, and strain-softening. This confirms that explicit friction is the primary driver of shear strength and structural integrity in CG clay models.

To the authors' knowledge, this study represents a pioneering effort to explicitly unify a granular contact force field with long-range molecular interactions within a single CGMD framework. By successfully coupling the physics of Hertzian contact mechanics with Buckingham potentials, this work provides a first-of-its-kind solution to the long-standing problem of unphysical fluid-like behavior in clay simulations. While this initial formulation serves as a foundational methodology, it establishes a robust new paradigm for clay modeling where macroscopic mechanical integrity is directly derived from a synergy of molecular energy landscapes and mesoscopic energy dissipation. This framework provides essential instructional guidance for future multi-scale research aiming to bridge the gap between atomistic descriptors and the complex, frictional reality of geotechnical systems.

## Acknowledgement

The authors gratefully acknowledge the financial supports from the Research Grants Council of Hong Kong (Project No. 15231825, N_PolyU534/20, 15217220).

## Data Availability Statement

All data that support the findings of this study are available from the corresponding author upon reasonable request.

## References

1. Wang, F., et al., *CTMAB-Modified Bentonite–Based PRB in Remediating Cr(VI) Contaminated Groundwater.* Water, Air, & Soil Pollution, 2020. **231**(1): p. 20.
2. Cui, L.-Y., C. Zhou, and W.-M. Ye, *Modified hydraulic conductivity equations of bentonite-based materials under saturated and unsaturated conditions.* Canadian Geotechnical Journal, 2025. **62**: p. 1-17.
3. Brochard, L., *Swelling of Montmorillonite from Molecular Simulations: Hydration Diagram and Confined Water Properties.* Journal of Physical Chemistry C, 2021. **125**(28): p. 15527-15543.
4. Al-Zaoari, K., et al., *Early stage of swelling process of dehydrated montmorillonite through molecular dynamics simulation.* Materials Chemistry and Physics, 2022. **283**: p. 126015.
5. Yuan, R., et al., *Molecular dynamics modelling of Na-montmorillonite subjected to uniaxial compression and unidirectional shearing.* Clay Minerals, 2022. **57**(3-4): p. 241-252.
6. Han, Z., et al., *A molecular dynamics study on the structural and mechanical properties of pyrophyllite and M-Montmorillonites (M = Na, K, Ca, and Ba).* Chemical Physics Letters, 2022. **803**.
7. Wei, P., et al., *Effect of water content and structural anisotropy on tensile mechanical properties of montmorillonite using molecular dynamics.* Applied Clay Science, 2022. **228**.
8. Zhou, A., et al., *Molecular modeling of clay minerals: A thirty-year journey and future perspectives.* Coordination Chemistry Reviews, 2025. **526**.
9. Noid, W.G., *Perspective: Coarse-grained models for biomolecular systems.* The Journal of Chemical Physics, 2013. **139**(9): p. 090901.
10. Ebrahimi, D., A.J. Whittle, and R.J.M. Pellenq, *Mesoscale properties of clay aggregates from potential of mean force representation of interactions between nanoplatelets.* Journal of Chemical Physics, 2014. **140**(15).
11. Ebrahimi, D., A. Whittle, and R.-M. Pellenq, *Effect of Polydispersity of Clay Platelets on the Aggregation And Mechanical Properties of Clay at the Mesoscale.* Clays and Clay Minerals, 2016. **64**(4): p. 425-437.
12. Ebrahimi, D., R.J.-M. Pellenq, and A.J. Whittle, *Mesoscale simulation of clay aggregate formation and mechanical properties.* Granular Matter, 2016. **18**(3).
13. Zhang, Y., et al., *A Coarse-Grained Interaction Model for Sodium Dominant Montmorillonite.* Langmuir, 2022. **38**(43): p. 13226-13237.
14. Zhang, Y., et al., *Mechanical properties and pore network connectivity of sodium montmorillonite as predicted by a coarse-grained molecular model.* Applied Clay Science, 2023. **243**.
15. Sun, H.-m., et al., *A coarse-grained water model for mesoscale simulation of clay-water interaction.* Journal of Molecular Liquids, 2020. **318**.

16. Ghazanfari, S., et al., *A Coarse-Grained Model for the Mechanical Behavior of Na-Montmorillonite Clay.* Langmuir, 2022. **38**(16): p. 4859-4869.
17. Zheng, X., X. Shen, and I.C. Bourg, *Coarse-grained simulation of colloidal self-assembly, cation exchange, and rheology in Na/Ca smectite clay gels.* J Colloid Interface Sci, 2025. **693**: p. 137573.
18. Potter, T.D., J. Tasche, and M.R. Wilson, *Assessing the transferability of common top-down and bottom-up coarse-grained molecular models for molecular mixtures.* Physical Chemistry Chemical Physics, 2019. **21**(4): p. 1912-1927.
19. Jin, J., et al., *Bottom-up Coarse-Graining: Principles and Perspectives.* Journal of Chemical Theory and Computation, 2022. **18**(10): p. 5759-5791.
20. Bandera, S., et al., *Effects of the Absence of Friction in Coarse-Grained Molecular Dynamics Simulations of Clay.* International Journal of Geomechanics, 2024. **24**(10).
21. Tang, Z.-Q., et al., *A novel mesoscale modelling method for steel fibre-reinforced concrete with the combined finite-discrete element method.* Cement and Concrete Composites, 2024. **149**: p. 105479.
22. Liu, D., Z.-Y. Yin, and C. O'Sullivan, *DEM Exploration of Stress Transmission and Small Strain Behavior of Rubber Sand Mixtures.* Journal of Geotechnical and Geoenvironmental Engineering, 2025. **151**(5): p. 04025031.
23. Shen, Z., et al., *DEM simulation of microscopic structure and macroscopic mechanical behavior of clay in oedometer and triaxial compression tests.* Computers and Geotechnics, 2024. **173**: p. 106544.
24. Liu, B., Z.-Y. Yin, and P.-Y. Hicher, *The compressibility of kaolin clay: a micromechanical perspective based on DEM.* Canadian Geotechnical Journal, 2026. **63**: p. 1-19.
25. Wang, Y. and N. Lu, *A Closed-Form Equation for Soil Sorptive Potential.* Journal of Geotechnical and Geoenvironmental Engineering, 2025. **151**(9).
26. Xu, W.-Q., Z.-Y. Yin, and Y.-Y. Zheng, *Investigating silica interface rate-dependent friction behavior under dry and lubricated conditions with molecular dynamics.* Acta Geotechnica, 2023. **18**: p. 3543–3554.
27. Thompson, A.P., et al., *LAMMPS - a flexible simulation tool for particle-based materials modeling at the atomic, meso, and continuum scales.* Computer Physics Communications, 2022. **271**: p. 108171.
28. Silbert, L.E., et al., *Granular flow down an inclined plane: Bagnold scaling and rheology.* Physical Review E, 2001. **64**(5): p. 051302.
29. Zheng, X. and I.C. Bourg, *Microstructure, Transport, and Mechanics of Compacted Clay Simulated at the 0.1 μm Scale (1400 Smectite Clay Particles) Using a Coarse-Grained Model with Explicit Counterions.* The Journal of Physical Chemistry C, 2026. **130**(10): p. 3990-4004.
30. Shen, X., X. Zheng, and I.C. Bourg, *A coarse-grained model of clay colloidal aggregation and consolidation with explicit representation of the electrical* *double layer.* J Colloid Interface Sci, 2025. **683**(Pt 1): p. 1188-1196.
31. Fan, J., et al., *Molecular dynamics predictions of thermomechanical properties of an epoxy thermosetting polymer.* Polymer, 2020. **196**.
32. Xu, W.Q., Z.Y. Yin, and Y.Y. Zheng, *FRP–soil interfacial mechanical properties with molecular dynamics simulations: Insights into friction and creep behavior.* International Journal for Numerical and Analytical Methods in Geomechanics, 2023(47): p. 2951-2967.
33. Wei, P., et al., *Nanoscale friction at the quartz-quartz/kaolinite interface.* Colloids and Surfaces A: Physicochemical and Engineering Aspects, 2023. **676**.
34. Wei, P., et al., *Nanoscale Stick-Slip Behavior and Hydration of Hydrated Illite Clay.* Computers and Geotechnics, 2024. **166**.
35. Morrow, C.A., D.E. Moore, and D.A. Lockner, *Frictional strength of wet and dry montmorillonite.* Journal of Geophysical Research: Solid Earth, 2017. **122**(5): p. 3392-3409.
36. Miari, M., K.K. Choong, and R. Jankowski, *Seismic pounding between adjacent buildings: Identification of parameters, soil interaction issues and mitigation measures.* Soil Dynamics and Earthquake Engineering, 2019. **121**: p. 135-150.
37. Baille, W., S. Tripathy, and T. Schanz, *Swelling pressures and one-dimensional compressibility behaviour of bentonite at large pressures.* Applied Clay Science, 2010. **48**(3): p. 324-333.
38. Marcial, D., P. Delage, and Y.J. Cui, *On the high stress compression of bentonites.* Canadian Geotechnical Journal, 2002. **39**(4): p. 812-820.
39. Tiwari, B. and B. Ajmera, *Consolidation and swelling behavior of major clay minerals and their mixtures.* Applied Clay Science, 2011. **54**(3-4): p. 264-273.
40. Cheng, G., et al., *Experimental Investigation of Consolidation Properties of Nano-Bentonite Mixed Clayey Soil.* Sustainability, 2020. **12**(2).
41. Chahal, R., et al., *Deep-Learning Interatomic Potential Connects Molecular Structural Ordering to the Macroscale Properties of Polyacrylonitrile.* ACS Applied Materials & Interfaces, 2024. **16**(28): p. 36878-36891.
42. Huang, W., et al., *Isomorphic α-Crystal and Amorphous Multiscale Structural Evolution Enables Heat-Resistant and High-Strength Semiaromatic Copolyamide Fibers.* Macromolecules, 2026. **59**(3): p. 1581-1590.
43. Zhu, H., et al., *Mesoscale simulation of aggregation of imogolite nanotubes from potential of mean force interactions.* Molecular Physics, 2019. **117**(22): p. 3445-3455.
44. Berzi, D., D. Vescovi, and B. Nadler, *Shaking into order: Q-tensor/kinetic theory of vibrated non-spherical grains in a confined geometry.* Journal of Fluid Mechanics, 2025. **1024**: p. A32.
45. Basconi, J.E. and M.R. Shirts, *Effects of Temperature Control Algorithms on Transport Properties and Kinetics in Molecular Dynamics Simulations.* Journal of Chemical Theory and Computation, 2013. **9**(7): p. 2887-2899.
46. Hicks, A., M. MacAinsh, and H.-X. Zhou, *Removing Thermostat Distortions of*

*Protein Dynamics in Constant-Temperature Molecular Dynamics Simulations.* Journal of Chemical Theory and Computation, 2021. **17**(9): p. 5920-5932.

47. Zhang, L.-L., et al., *Nanoscale mechanical behavior of kaolinite under uniaxial strain conditions.* Appl Clay Sci, 2021. **201**.
48. Xiao, C., et al., *Mechanical properties of defective kaolinite in tension and compression: A molecular dynamics study.* Applied Clay Science, 2023. **246**: p. 107164.
49. Thapa, K.B., K.S. Katti, and D.R. Katti, *Compression of Na–Montmorillonite Swelling Clay Interlayer Is Influenced by Fluid Polarity: A Steered Molecular Dynamics Study.* Langmuir, 2020. **36**(40): p. 11742-11753.
50. Faisal, H.M.N., K.S. Katti, and D.R. Katti, *Molecular mechanics of the swelling clay tactoid under compression, tension and shear.* Applied Clay Science, 2021. **200**.
51. Mondol, N.H., et al., *Experimental mechanical compaction of clay mineral aggregates—Changes in physical properties of mudstones during burial.* Marine and Petroleum Geology, 2007. **24**(5): p. 289-311.
52. Anglade, E., et al., *Physical and mechanical properties of clay–sand mixes to assess the performance of earth construction materials.* Journal of Building Engineering, 2022. **51**: p. 104229.
53. Li Xue, P., et al., *Compressive Modulus of Soil-Bentonite Mixtures for Cutoff Walls from CPTU Data*, in *Geo-Chicago 2016*. 2016. p. 528-536.

# Tables

Table 1. Summary of Hybrid Potential Parameters for CG-MMT Interactions.

| Pair | Buckingham | | Hertz | Diameter (Å) |
|---|---|---|---|---|
| | A (kcal/mol) | $\rho$ (Å) | $k_n$ (kcal/(mol·Å$^{3/2}$)) | |
| Inner-Inner | 1200 | 0.78 | 0.2 | 8.5 |
| Outer-Outer | 600 | 0.70 | 0.08 | |
| Inner-Outer | 400 | 1.2 | 1.3 | |

Table 2. Summary of CG-MMT structural properties after isotropic compression under varying damping parameters.

| Parameter Set | Order Parameter $S$ | System Director **n** | Void ratio $e$ |
|---|---|---|---|
| $x_{\gamma,t} = 5$ | | | |
| $\eta_n = 0.2$ | 0.86 | [-0.025, -0.053, 0.998] | 1.03 |
| $\eta_n = 1.0$ | 0.18 | [0.542, 0.807, 0.234] | 1.45 |
| $\eta_n = 2.0$ | 0.04 | [0.035, -0.660, 0.751] | 1.97 |
| $\eta_n = 0.2$ | | | |
| $x_{\gamma,t} = 1$ | 0.80 | [-0.997, 0.048, 0.067] | 1.06 |
| $x_{\gamma,t} = 10$ | 0.37 | [-0.963, 0.156, 0.219] | 1.23 |